\documentclass[a4paper,11pt]{article}

\usepackage{jinstpub} 

\usepackage{lineno}
\usepackage{siunitx}
\usepackage{float}
\usepackage{caption}
\usepackage{subcaption}
\usepackage{xcolor}

\newcommand{\PN}{\emph{PoolNet}}
\newcommand{\WN}{\emph{WideNet}}
\newcommand{\DN}{\emph{DeepPoolNet}}
\DeclareSIUnit\photons{photons}

\title{\boldmath Partition Pooling for Convolutional Graph Network Applications in Particle Physics}

\author{M.~Bachlechner,}
\author[1]{T.~Birkenfeld\note{Corresponding author.},}
\author{P.~Soldin,}
\author{A.~Stahl}
\author{and C.~Wiebusch}

\affiliation{III Physics Institute B, RWTH Aachen University,\\
Otto-Blumenthal-Straße, Aachen, Germany}

\emailAdd{birkenfeld@physik.rwth-aachen.de}

\abstract{Convolutional graph networks are used in particle physics for effective event reconstructions and classifications.
However, their performances can be limited by the considerable amount of sensors used in modern particle detectors if applied to sensor-level data.
We present a pooling scheme that uses partitioning to create pooling kernels on graphs, similar to pooling on images.
Partition pooling can be used to adopt successful image recognition architectures for graph neural network applications in particle physics.
The reduced computational resources allow for deeper networks and more extensive hyperparameter optimizations.
To show its applicability, we construct a convolutional graph network with partition pooling that reconstructs simulated interaction vertices for an idealized neutrino detector.
The pooling network yields improved performance and is less susceptible to overfitting than a similar network without pooling.
The lower resource requirements allow the construction of a deeper network with further improved performance.}

\keywords{Analysis and statistical methods; Data processing methods; Large detector-systems performance; Pattern recognition, cluster finding, calibration and fitting methods}

\arxivnumber{2208.05952} 

\begin{document}
\maketitle
\flushbottom

\section{Introduction}

    Convolutional Neural Networks (CNNs) are widely used in particle physics for numerous reconstruction tasks \cite{Feickert:2021ajf}.
    Their basic structure of trainable convolutions followed by pooling accomplishes pattern recognition, resulting in effective parameter estimations or event classifications \cite{Feickert:2021ajf, CNN_goodfellow, CNN_geron2019hands}.
    Classical CNNs have been developed for image processing tasks.
    Images are organized as two-dimensional, regular grids of pixels, with color information as an additional feature dimension.
    Events recorded with particle physics detectors resemble images, but are often not arranged in such a two-dimensional grid.
    Experimental requirements lead to different detector geometries and sensor arrangements, e.g., cylindrical \cite{DoubleChooz:2006vya}, hexagonal \cite{IceCube:2016zyt}, or spherical \cite{JUNO:2015sjr}.
    A projection or transformation of the sensor positions onto a regular grid is generally required to apply a classical CNN to sensor-level data.
    A geometric transformation can introduce unwanted side effects \cite{projections}:
    Distortions of distances, areas, volumes, or angles are inevitable for transformations such as, e.g., Mercator or Hammer-Aitoff projections \cite{mercator,hammer_aitoff}. 
    
    In recent years, graph neural networks (GNNs) have been used to overcome this issue \cite{GEOM_DL, GNNs, Shlomi:2020gdn, Qasim:2019otl}.
    A \emph{graph} is a data structure consisting of nodes and selected connections between nodes, called edges \cite{graph_theory}. 
    Nodes typically contain data, e.g., of one individual sensor.
    Edges encode information about the spatial correlation between associate nodes \cite{graph_signal}.
    A GNN can learn representations of the data under given spatial correlations.
    By generalizing convolutions to graphs, their feature extraction can be utilized directly on the detector’s natural geometry \cite{GEOM_DL, GNNs, GCN}.
    Transformation to a regular grid is no longer required.
    We refer to networks containing convolution layers on graphs as \emph{Convolutional Graph Networks} (CGNs) in this work.
      
    For modern detectors with a considerable number of sensors, convolutions may require a significant computational effort, especially for deep networks.
    Pooling lowers the required computational resources and improves the effectiveness of training by reducing the dimensionality of the hidden layers successively \cite{CNN_goodfellow, CNN_geron2019hands}.
    It makes deeper networks accessible, as resources are limited in real-life applications.
    In conventional CNNs, pooling also ensures translational invariance of feature recognition \cite{Original, CNN_geron2019hands}.
    With CGN pooling similar to CNN pooling, we can adopt successful CNN architectures to applications on arbitrary geometries.

    We present a pooling scheme that utilizes \emph{graph partitioning} \cite{graph_partition}.
    Graph partitioning is a common tool to dissect graphs into non-overlapping groups of adjacent nodes, called partitions.
    It can be used for arbitrary graphs and does not rely on case specific sensor patterns or symmetries.
    We adopt a typical CNN architecture to a CGN using \emph{partition pooling} and test it on the use case of a typical particle detector.
    Even though our test scenario is strongly motivated by modern neutrino physics, CGNs utilizing partition pooling can be applied to any detector with any geometric sensor arrangement.
    As presented in section \ref{sec:pooling}, partition pooling solely relies on the spatial position of the sensors.
    The method can be evaluated for any reconstruction, e.g., energy and track reconstruction, and classification tasks.

\section{Detector Geometries in Graph Representation}
\label{sec:graphs}
   
    A \emph{graph} $\mathcal{G}$ is a structure that stores data in nodes which are connected via edges \cite{graph_theory, graph_signal}.
    We can describe it via a pair $\mathcal{G} = (\mathbf{N}, \mathbf{A})$ consisting of the node vector $\mathbf{N} \in \mathbb{R}^{n \times d}$ where $n \in \mathbb{N}$ is the number of nodes and $d \in \mathbb{N}$ the amount of data at each node, and the adjacency matrix $\mathbf{A} \in \mathbb{R}^{n \times n}$ with
    \begin{equation}
        A_{ij} = 
            \begin{cases}
            1,              & \text{if node $i$ and $j$ connected}\\
            0,              & \text{otherwise}
            \end{cases}
        \qquad .
    \end{equation}
    
    We can adapt this formalism to describe the data measured by particle detectors.
    A detector typically consists of sensors or groups of such, $\mathbf{s}$, that measure multiple observables.
    The position of sensor $s_i$ $(i=1 \ldots n)$ is given as $\mathbf{r}_i$.
    We assign a node to each sensor.
    Node entry $\mathbf{N}_i$ contains the data of sensor $s_i$ as features $N_{i,\,f} \: (f=1 \ldots d)$.
    We weight $A_{ij}$ depending on the distance between the corresponding sensors to encode the detector's geometry.
    The weighted adjacency matrix, that we use for the application of graph convolutions, is defined as 
    \begin{equation}
        \label{eq:adj}
        W_{ij} = \frac{A_{ij}}{1+ \left( \frac{\Delta_{ij}}{\Delta_0} \right)^2}, 
    \end{equation}
    with the Euclidean distance $\Delta_{ij} = | \mathbf{r}_i - \mathbf{r}_j |$ and $\Delta_0$ the normalization scale of distances in the graph.
    We use the Euclidean distances in our example because this metric can be used universally for all detectors.
    We only connect sensors that are close to each other (we refer to section \ref{sec:ana} for details).
    Connecting only geometrically neighboring nodes restricts the convolutions to filter locally confined features \cite{GCN, cheby}.

\section{Pooling via Graph Partitioning}
\label{sec:pooling}
    In conventional CNNs a constant pooling kernel can stride over an image \cite{CNN_goodfellow, CNN_geron2019hands}.
    This method can not be transferred to CGNs in general, as the local graph structure can vary.
    However, every graph can be dissected by partitioning \cite{graph_partition}.
    Partitioning algorithms group close-by nodes to $M$ approximately equal-sized sub-graphs.
    We can describe the grouping as
    \begin{equation}
        \mathbf{P} = \left( P_1, P_2, ~ \dots, \, P_M \right),
    \end{equation}
    for which $P_k$ $(k=1 \ldots M)$ is one group of nodes. 
    We refer to $P_k$ as a partition of the partitioning $\mathbf{P}$.
    The partitions are adjacent and non-overlapping, as each node is part of exactly one partition.
    We use the \emph{K-means} algorithm \cite{k_means} to dissect the graphs.
    Other partitioning algorithms might be equally suited.
    Using the resulting partitions as kernels, we can implement a pooling.
    We apply an aggregation function $H$ over each partition $P_k$ to all feature spaces $f$
    \begin{equation}
        N_{k,\,f}' = \underset{\forall i \, \in \, P_k}{H} (N_{i,\,f}) \, .
    \end{equation}
    $H$ can be any aggregation function.
    We use $H(x) = \max(x)$ for our example, as it often yields better performances in feature recognition tasks on images compared to, e.g., average pooling \cite{hmax_pool}.
    Applying $H$ over the partitions is similar to the conventional usage of pooling in CNNs, as the partitions are adjacent and non-overlapping.
    
\begin{figure}[h]
        \centering
        \includegraphics[width=0.99\textwidth]{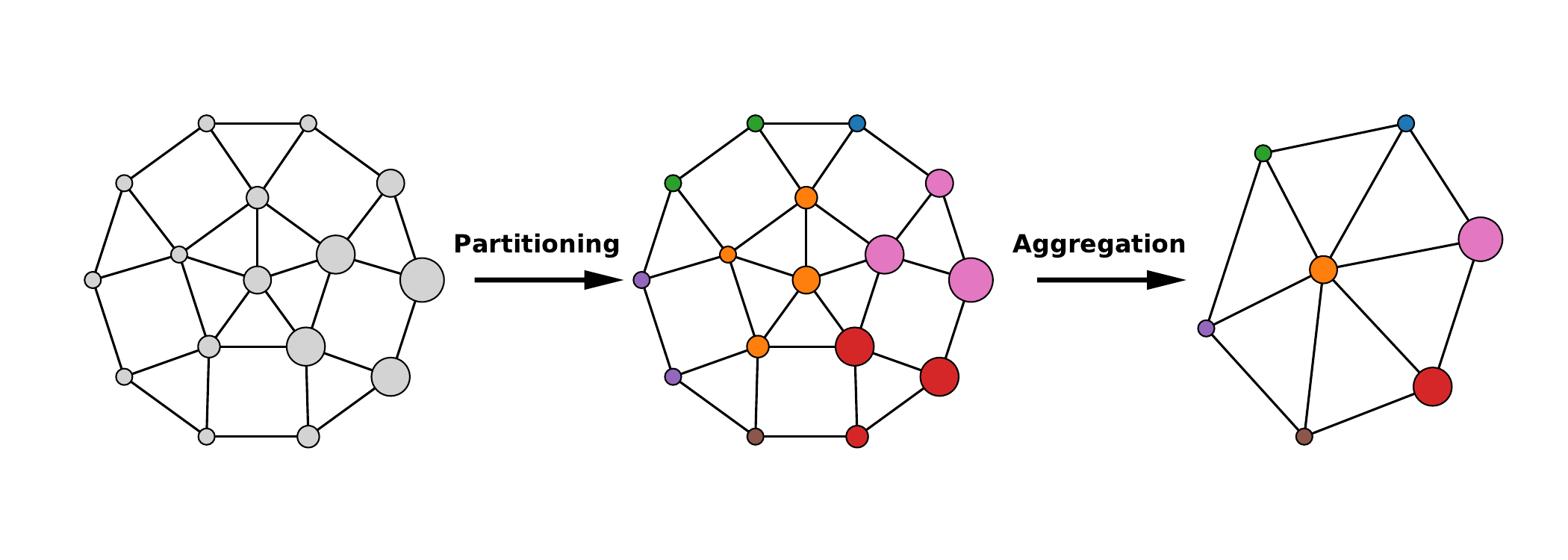}
        \caption{Scheme of the partition pooling. An example graph is partitioned via the \emph{K-Means} algorithm. Nodes of the same partition are depicted with the same color. An example signal over the graph is shown by enlarging the nodes proportionally to the signal strength. A $\max$-aggregation is applied over the partitions, and the output is shown in the resulting graph structure.
        The example graph has a similar structure as the detector lids in section \ref{sec:ana}. The pooling is applied similarly.}
        \label{fig:pool_sketch}
    \end{figure}

    The aggregation results $N_{k,\,f}'$ are generally forwarded to a succeeding CGN layer.
    Therefore, we represent the results in a new graph $\mathcal{G}'$ as follows:
    Each $N_{k,\,f}'$ is regarded as feature $f$ of a new node $\mathbf{N}_k'$.
    The average position of the nodes in $P_k$ is assigned to the new node as its position
    \begin{equation}
        \mathbf{r}_k' =  \frac{1}{n_k} ~ \sum\limits_{~ i \, \in \, P_k} ~ \mathbf{r}_i ,
    \end{equation}
    with $n_k$ being the number of nodes in $P_k$.
    Thus the node structure of $\mathcal{G}'$ is static with respect to the individual sample the network is applied to and does not change dynamically, similar to the fix pixel grids resulting from pooling on images.
    We construct the adjacency matrices $\mathbf{A}'$ and $\mathbf{W}'$ of $\mathcal{G}'$ from the new positions according to equation \ref{eq:adj}, 
    \begin{equation}
        \label{eq:new_adj}
        A_{kl}' =
        \begin{cases}
        1,         & \text{if new node $k$ and $l$ connected}\\
        0,              & \text{otherwise}
        \end{cases} \, , \quad
        W_{kl}' = \frac{A_{kl}'}{1+ \left( \frac{\Delta_{kl}'}{\Delta_0} \right)^2}, 
    \end{equation}
    with $\Delta_{kl}' = | \mathbf{r}_k' - \mathbf{r}_l' |$.
    Similar to the original detector graph, we connect only close-by nodes for the application of convolutions.
    A sketch of the pooling is shown in Figure \ref{fig:pool_sketch}.
 
    As we have a full description of $\mathcal{G}'$, including node vector $\mathbf{N}'$, adjacency matrices $\mathbf{A}', \mathbf{W}'$, and node positions $\mathbf{r}_j'$, we can repeatedly apply the pooling operation, similar to pooling in CNNs
    \begin{equation}
        \mathcal{G} \overset{H-\mathrm{pool}}{\longrightarrow} \mathcal{G}' \overset{H'-\mathrm{pool}}{\longrightarrow} \mathcal{G}'' \overset{H''-\mathrm{pool}}{\longrightarrow} ~ \dots \quad .
    \end{equation}
    
    Under the assumption that the detector geometry is constant, the pooled graph structures are the same for all events in the data set.
    The partitioning and calculation of the new adjacency matrices can be done once, prior to training, for a given set of hyperparameters.
    The application of the aggregation function $H$ is the only operation that needs to be computed during runtime, but this can be implemented efficiently.
    
    We have implemented the partition pooling using \emph{tensorflow} \cite{tensorflow}.
    It computes the aggregation for a given partitioning. 
    The partitioning needs to be provided by the user.
    \emph{tensorflow} does not provide a CGN implementation.
    An extension is required, e.g., \emph{spektral} \cite{spektral}.
    The source code of the partition pooling is available at \cite{git_repo}.

\section{An Example Pooling Application
\label{sec:ana}}

    For demonstration, we apply our pooling scheme in a CGN that adopts a typical CNN architecture.
    We compare it to a similar CGN without pooling.
    For that purpose, we have created a simulation of a simplified neutrino detector.
    We train and apply the CGNs to reconstruct the neutrino interaction vertices.
    For \SI{}{\MeV} neutrinos, e.g., emitted by nuclear reactors, the neutrino target is usually a liquid scintillator vessel surrounded by a considerable large number of photomultiplier tubes (PMTs) \cite{reactor_neutrino_book, reactor_neutrino_paper}. 
    The vessel typically leads to a curved PMT arrangement, making a detector like this an ideal test scenario for partition pooling.
    
    \subsection{The Simulation}
    
\begin{figure}[ht]
        \centering
        \includegraphics[width=0.99\textwidth]{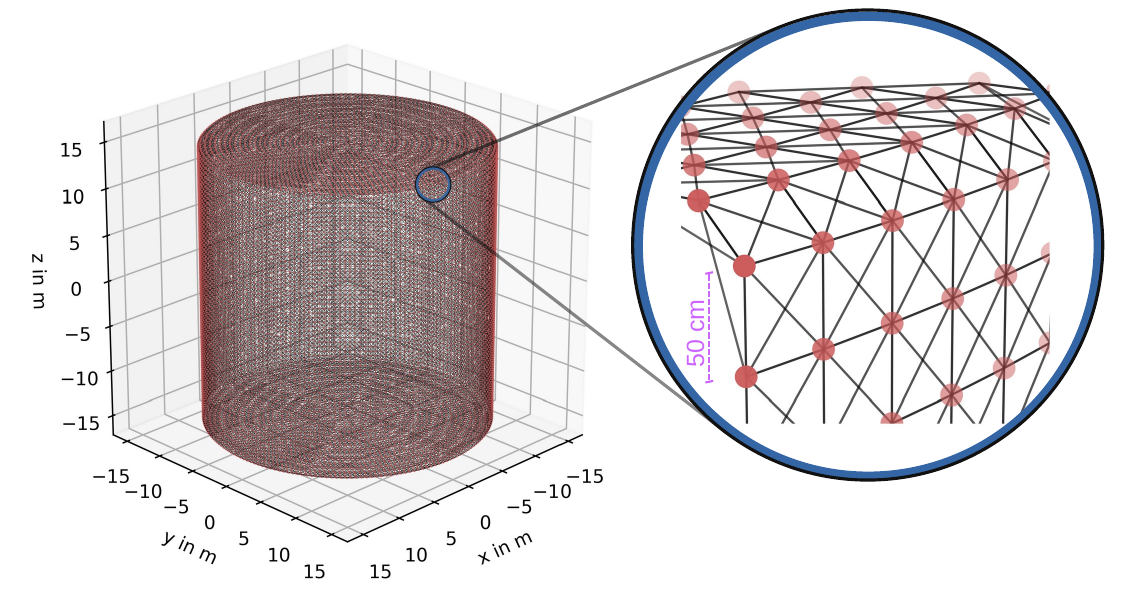}
        \caption{Detector graph. The nodes, associated with the PMTs, are shown as red dots. The edges are shown in black. For visualization, only the eight closest neighbors are connected. The number of connections used in our CGN implementation varies, we refer to section \ref{sec:reco} for details.}
        \label{fig:toy_graph}
    \end{figure}

\begin{figure}[ht]
        \centering
        \includegraphics[width=0.75\textwidth]{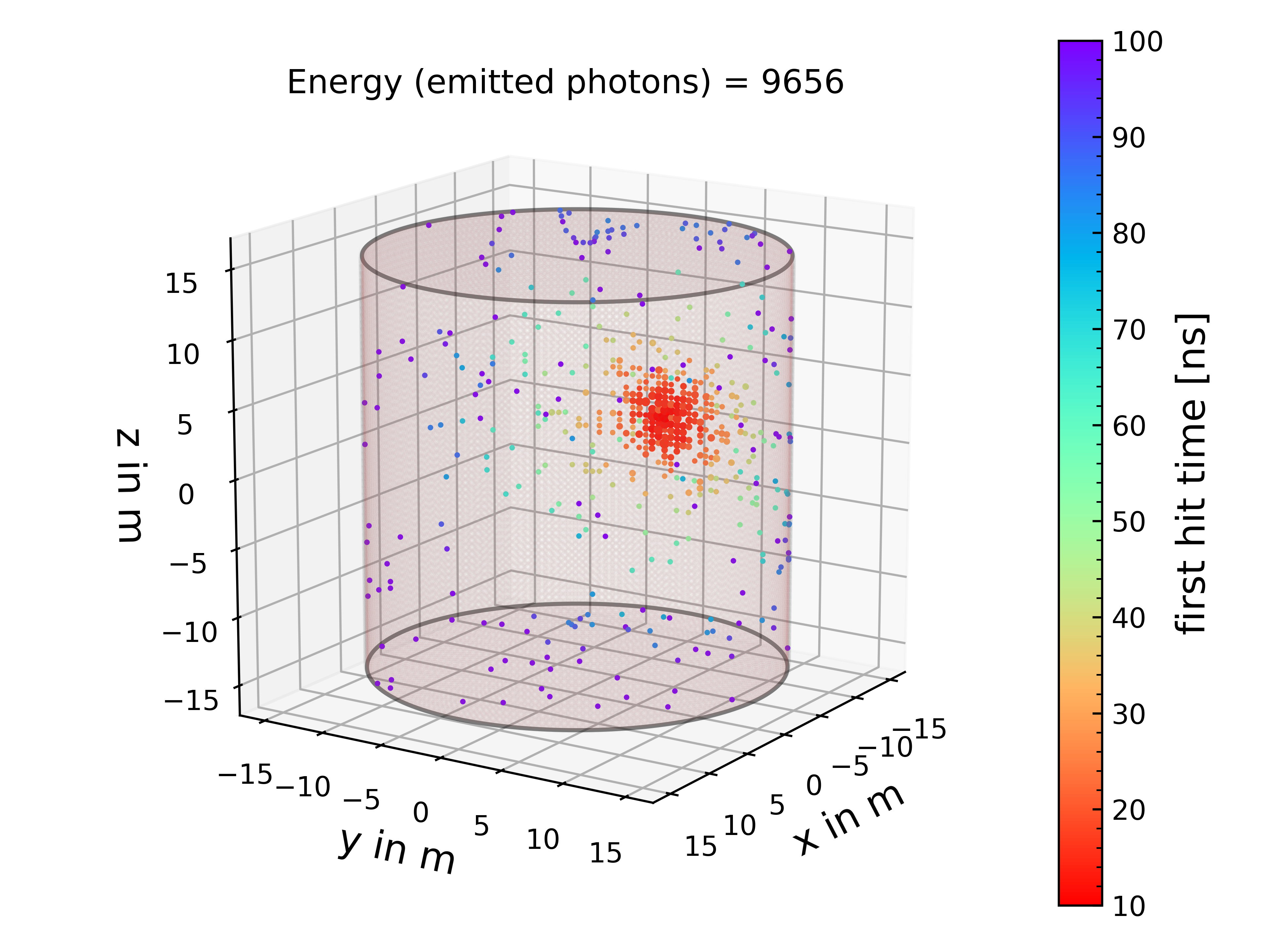}
        \caption{Event display: The geometry of the detector setup is shown. Each fired PMT is represented by a colored dot. The color denotes the arrival time of the first registered photon. The size of the dot is proportional to the total number  of detected photons. The vertex of this event is close to the rear wall of the detector. Therefore a cluster of multiple hit PMTs with early arrival times can be seen  close to it.}
        \label{fig:event_sample}
    \end{figure}

    Our detector consists of a cylindrical liquid scintillator target with a radius of \SI{15}{\meter} and height of \SI{30}{\meter}.
    The scintillation photons are detected by \num{16904} inward-facing PMTs situated on the detector surface.
    We include typical detector effects like the time profile of the scintillation, photon absorption, photon detection efficiency, and photon arrival time resolution. 
    For a more detailed description of the simulation, we refer to appendix \ref{apd:toy_mc}.
    The detector graph is depicted in Figure~\ref{fig:toy_graph}.
    A data set of \num{e6} point-like events has been generated, distributed uniformly over the target volume.
    The number of initial scintillation photons is uniformly distributed between \num{e3} and \num{e4}.
    An exemplary event is displayed in Figure \ref{fig:event_sample}.
    
    \subsection{Vertex Reconstruction with a Convolutional Graph Network}
    \label{sec:reco}
    We parameterize the vertex position in Cartesian coordinates $\mathbf{r}_{\mathrm{vtx}} = (x, y, z)$ and implement its reconstruction as regression which we optimize by supervised learning.
    We reduce the individual PMT data to the measured arrival time of the first photon $t_{i}^{\mathrm{first}}$ and the total number of detected photons $n^{\gamma}_i$, and feed both as inputs at the corresponding node
    \begin{equation}
        \mathbf{N}_{\mathrm{input}} = \begin{pmatrix}
                            t^{\mathrm{first}}_1 && n^{\gamma}_1 \\
                            t^{\mathrm{first}}_2 && n^{\gamma}_2 \\
                            \vdots && \vdots \\
                            t^{\mathrm{first}}_{\SI{16904}{}} && n^{\gamma}_{\SI{16904}{}}
                         \end{pmatrix} \, .
    \end{equation}
    
    If a PMT does not detect a photon, we set $t_{i}^{\mathrm{first}} = \SI{-10}{\nano \second}$, which is not ambiguous with physical arrival times which occur later.
    Both, $t_{i}^{\mathrm{first}}$ and $n^{\gamma}_i$ are sensitive to the interaction vertex as $t^{\mathrm{first}}_i \approx t^{\mathrm{ToF}}_i \propto r$ and  $n^{\gamma}_i \propto \frac{1}{r^2}$, with $t^{\mathrm{ToF}}_i$ the time of flight of the photon and $r = \left| \mathbf{r}_{\mathrm{PMT}} - \mathbf{r}_{\mathrm{vtx}} \right|$ the distance from the vertex to the PMT position $\mathbf{r}_{\mathrm{PMT}}$.
    
\begin{figure}[htbp]
        \centering
        \includegraphics[clip, trim=3.5cm 3.5cm 1.5cm 3.5cm, width=0.95\textwidth]{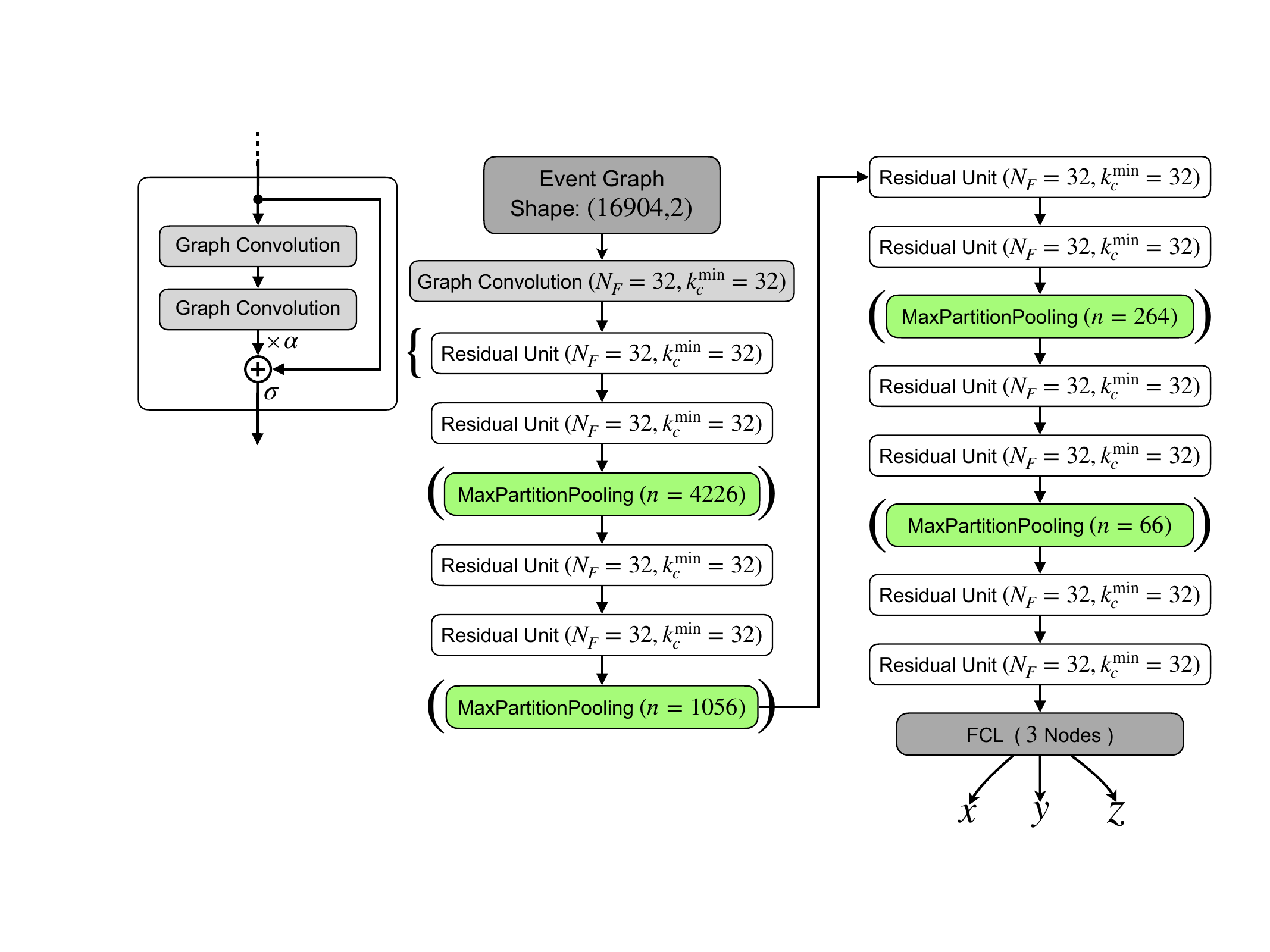}
        \caption{Network structures for \PN{} and \WN. The networks are fed with the node input. Each first layer is a single graph convolution layer. All following convolution layers are arranged in ten \emph{Residual} units \cite{ResNet} with additional residual weights \cite{ReZero}. For each convolution layer, the number of feature maps $N_F$, and the minimum connections per node $k_C^{\mathrm{min}}$ are given. After every second \emph{Residual} unit, a partition pooling layer, highlighted in green, reduces the number of nodes $n$ in \PN{} by about a factor of four. The networks outputs are the predicted three vertex coordinates $x,y,z$, implemented as fully connected layers with linear activations.}
        \label{fig:architecture}
    \end{figure}

    We base our networks on a sufficiently complex network architecture which has been shown to be suitable for vertex reconstructions in DoubleChooz \cite{markus_thesis}.
    As we show in section \ref{sec:train}, the architecture performs reasonably well for our setup.
    It is deep enough to demonstrate the benefits of multiple partition pooling stages.
    Finally, the architecture is comparable to typical network architectures for image recognition tasks.
    Thus we expect that a similar pooling will improve the network performance like-wise.
    We use a network with pooling which we call \PN, and a baseline network without pooling which we call \WN. 
    Both networks are illustrated in Figure~\ref{fig:architecture}.
    Each consists of twenty-one \emph{graph convolution} layers (GCN) in total, as introduced by Kipf and Welling \cite{GCN} with \emph{ReLU} activation \cite{CNN_goodfellow, CNN_geron2019hands}.
    We use the CGN implementation of \emph{spektral} \cite{spektral}, modify the restriction on the Chebyshev coefficients to $\theta'_0 = \theta'_1$, compare with equation 6 in \cite{GCN}, and omit the renormalization (equation~8 in \cite{GCN}).
    These adjustments lead to slighlty improved performances in our case.\footnote{The involved repetition of matrix multiplications does not necessarily suffer from numerical instabilities as the maximum eigenvalue of the corresponding matrix tends towards one for large graphs.}
    The first layer is a single GCN layer.
    All other GCN layers are combined to \emph{Residual Units} of two layers each, which we adopt from \cite{ResNet} and add residual weights $\alpha$ similar to \emph{ReZero Units} \cite{ReZero}.
    We use $N_F = \SI{32}{}$ feature maps uniformly.
    The amount of feature maps is limited by the computational resources required for \WN.
    As explained in section \ref{sec:graphs}, the effective kernel size is given by the number of connections per node within the graph structure fed to the GCN layer.
    We aim to connect a node only to its $k_C = \SI{32}{}$ closest neighboring nodes.
    However, a node is not necessarily one of the $k_C$'th closest neighbors of all its closest neighboring nodes in return. 
    Using this simple criterion would result in a directed graph.
    The GCN implementation requires an undirected graph.
    We connect a node to its $k_C = \SI{32}{}$ closest neighbors and to the remaining nodes for which the node is one of their closest neighbors.
    Hence, each node is connected to at least $k_C^{\mathrm{min}} = \num{32}$ other nodes.
    We set the scaling parameter $\Delta_0$ of the weighted adjacency matrices (equations \ref{eq:adj} and \ref{eq:new_adj}) to \SI{1}{\meter}.
    The final layer consists of three fully connected neurons with linear activation, each corresponding to one of the Cartesian coordinates.
        
\begin{figure}
    \centering
        
        \begin{subfigure}{.45\textwidth}
        \centering
        \includegraphics[clip, trim=14cm 11cm 10cm 11cm, width=.99\linewidth]{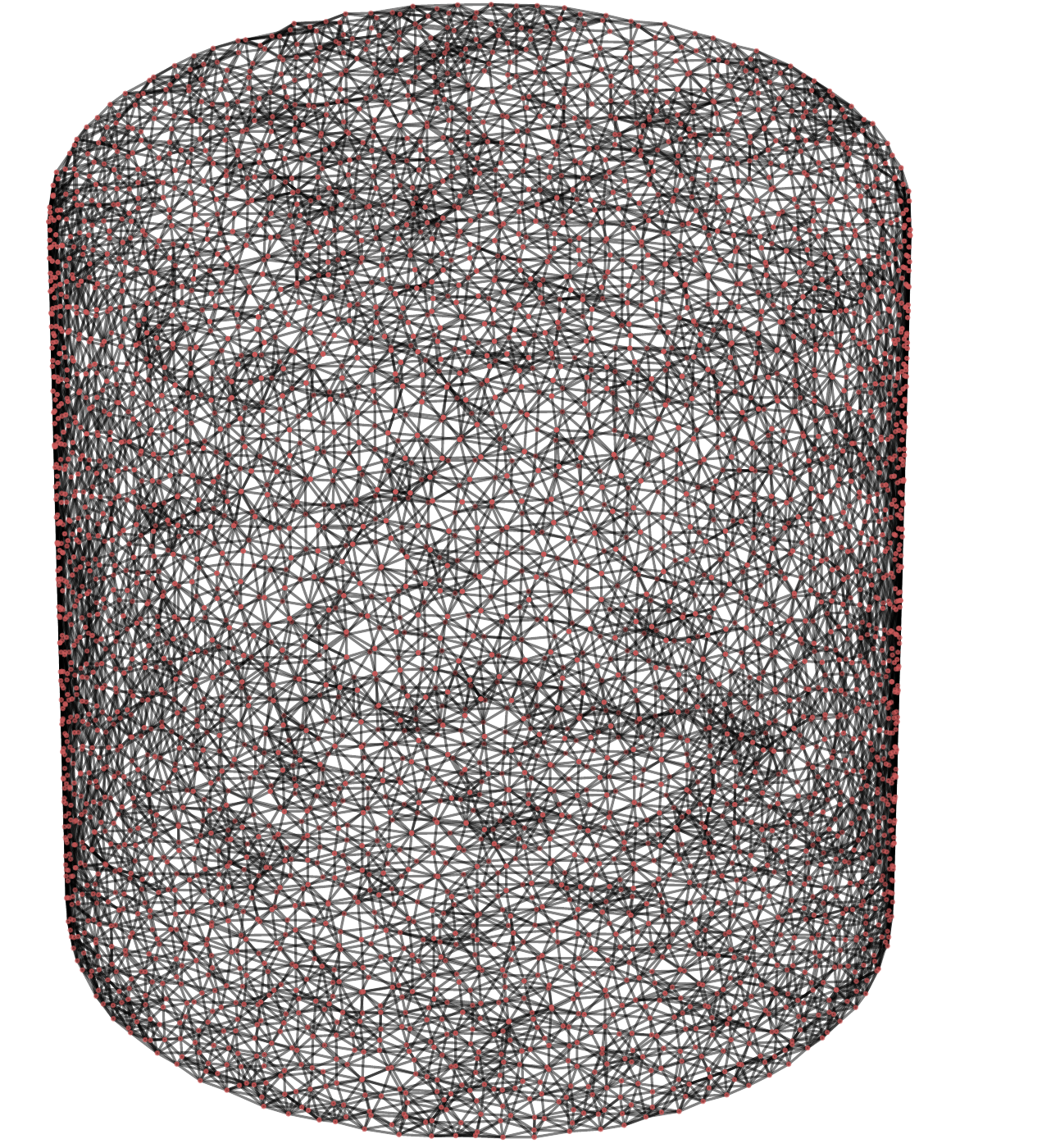}
        \caption{~ 1st Pooling: ~\SI{4226} nodes}
        \label{fig:pool_step1}
        \end{subfigure}
        \begin{subfigure}{.45\textwidth}
        \centering
        \includegraphics[clip, trim=14cm 11cm 10cm 11cm, width=.99\linewidth]{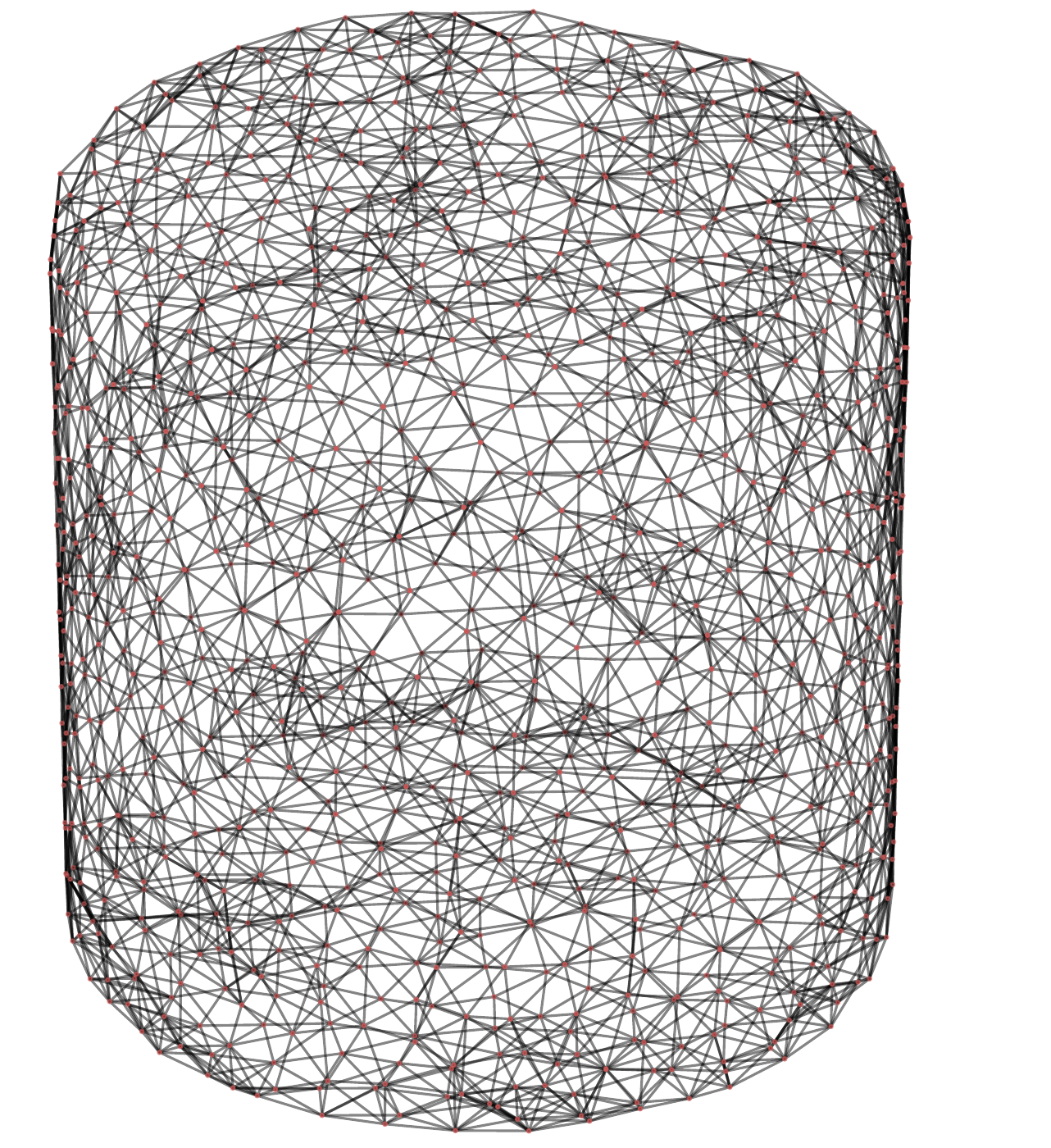}
        \caption{~ 2nd Pooling: ~\SI{1056} nodes}
        \label{fig:pool_step2}
        \end{subfigure}
        
        \begin{subfigure}{.45\textwidth}
        \centering
        \includegraphics[clip, trim=14cm 11cm 10cm 11cm, width=.99\linewidth]{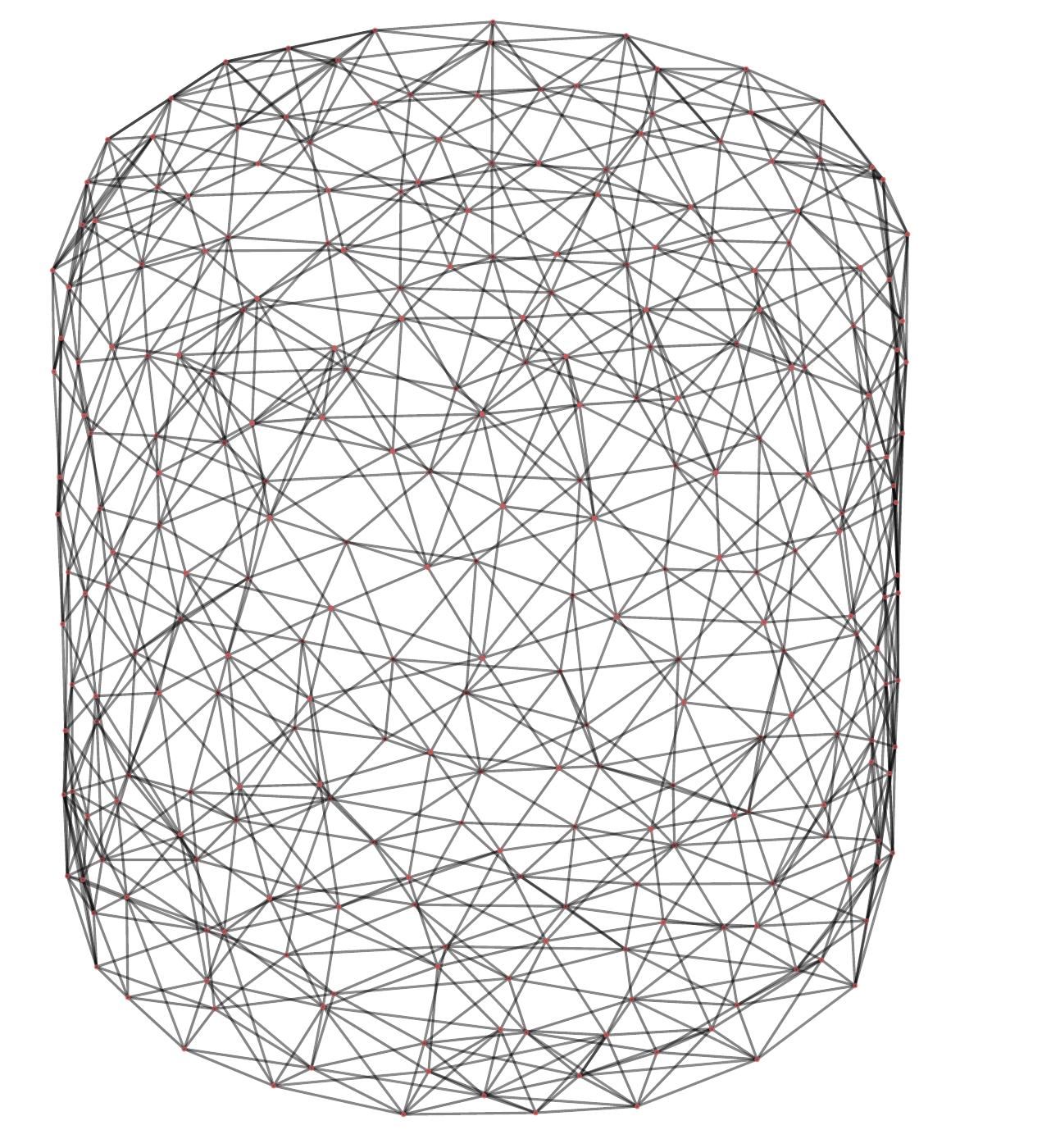}
        \caption{~ 3rd Pooling: ~\SI{264} nodes}
        \label{fig:pool_step3}
        \end{subfigure}
        \begin{subfigure}{.45\textwidth}
        \centering
        \includegraphics[clip, trim=14cm 11cm 10cm 11cm, width=.99\linewidth]{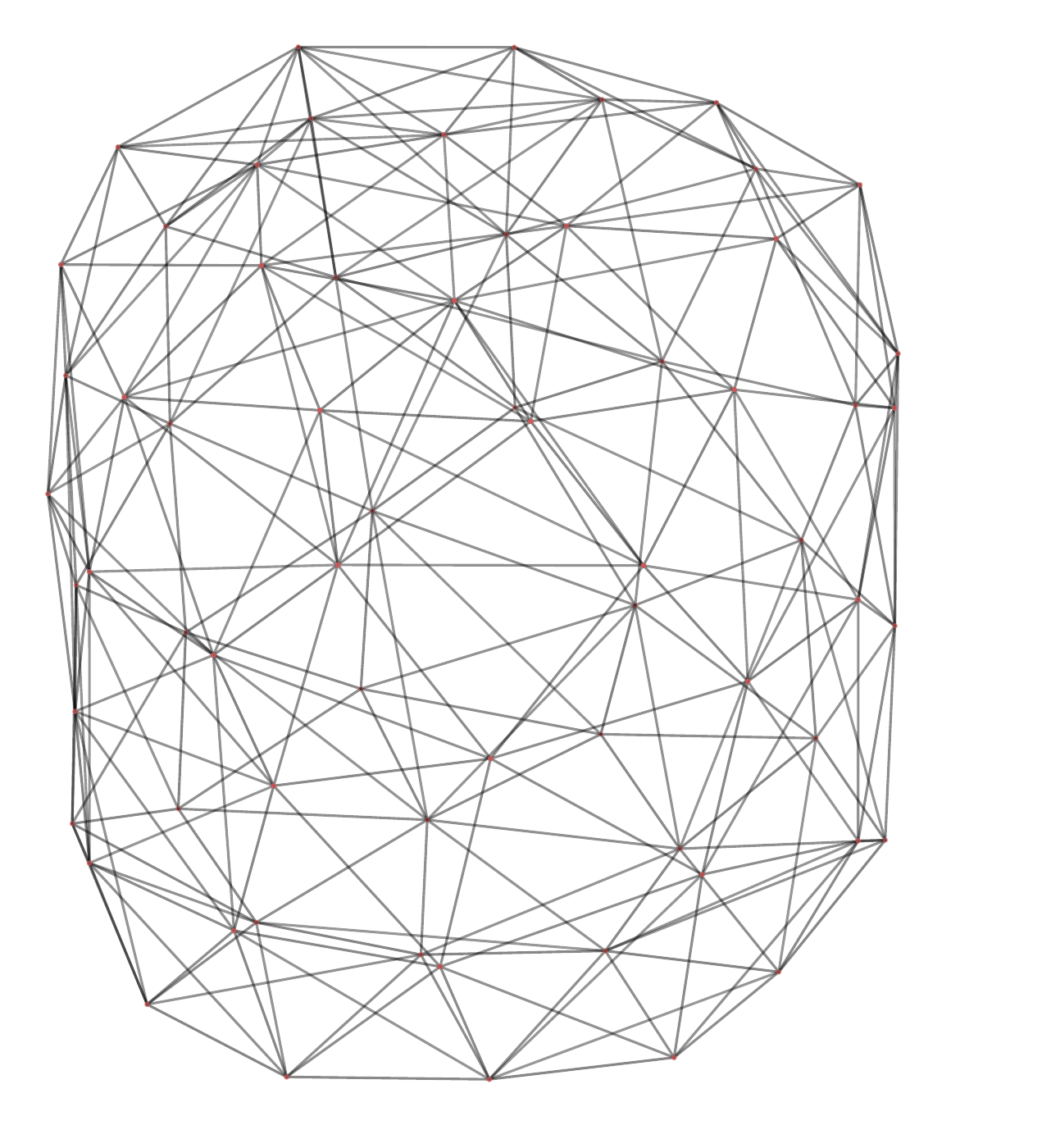}
        \caption{~ 4th Pooling: ~\SI{66} nodes}
        \label{fig:pool_step4}
        \end{subfigure}
        
        \caption{Graph structures after partition pooling.
        Each step reduces the graph size by roughly a factor of four.}
        \label{fig:pool_steps}
    \end{figure}

    The four pooling layers, surrounded by brackets and highlighted in green in Figure \ref{fig:architecture}, are only realized in \PN.
    Each pooling layer reduces the amount of nodes by about a factor of four, analogously to a $\SI{2}{} \times \SI{2}{}$ image pooling \cite{CNN_goodfellow, CNN_geron2019hands}.
    The resulting graph structures are shown in Figure~\ref{fig:pool_steps}.
    \PN{} and \WN{} are equivalent, except for the pooling layers. 
    We do not conduct  further fine-tuning of the networks, e.g., an in-depth hyperparameter optimization.
    The values of $k_C^{\mathrm{min}}$, $\Delta_0$ and the number of feature maps are empirically found to yield good results.
    
    The networks are trained using the \emph{ADAM} algorithm \cite{Kingma:2014vow}.
    We use the Mean Squared Error (MSE) as loss function, which becomes
    \begin{equation}
        \mathrm{MSE} = 
        \frac{1}{N_{\mathrm{batch}}} \sum\limits_{\mathrm{batches}} ~ \frac{1}{3} \left[ \left(x^{\mathrm{pred.}} - x^{\mathrm{truth}} \right)^2  + \left(y^{\mathrm{pred.}} - y^{\mathrm{truth}} \right)^2 + \left(z^{\mathrm{pred.}} - z^{\mathrm{truth}} \right)^2\right]
    \end{equation}
    for the three coordinates $x$, $y$, $z$, with $N_{\mathrm{batch}}$ being the number of events per batch.
    The superscript \emph{pred.} indicates the coordinate prediction of the network and \emph{truth} the true value sampled by the simulation.
    MSE is an estimator for the coordinate-averaged variance for an unbiased prediction.
    The data is split into the training set, consisting of \num{800000} events, and the validation and test sets, each consisting of \num{100000}.

    The total number of trainable parameters in the GCN layer set is \num{20906} for both networks.
    \WN{} has \num{1622787} trainable parameters in the last dense layer, whereas \PN{} has \num{6339} only as the pooling reduces the dimensionality.
    Overfitting is expected for such a large number of parameters in \WN{} \cite{CNN_goodfellow, CNN_geron2019hands}, while the smaller number of parameters in \PN{} makes the network less susceptible to it.
    In \PN, the convolutional layers account for a larger fraction of trainable parameters than in \WN.
    Thus, we expect that the convolutional layers in \PN{} are more relevant than in \WN.
    To show the different performances, we first train both networks without any additional regularization, e.g., \emph{dropout} \cite{dropout} (\emph{dropout} ratio of last layer $p_d = 0$).
     
    \subsection{Training and Evaluation \label{sec:train}}
\begin{figure}[htbp]
        \centering
        \includegraphics[width=0.75\textwidth]{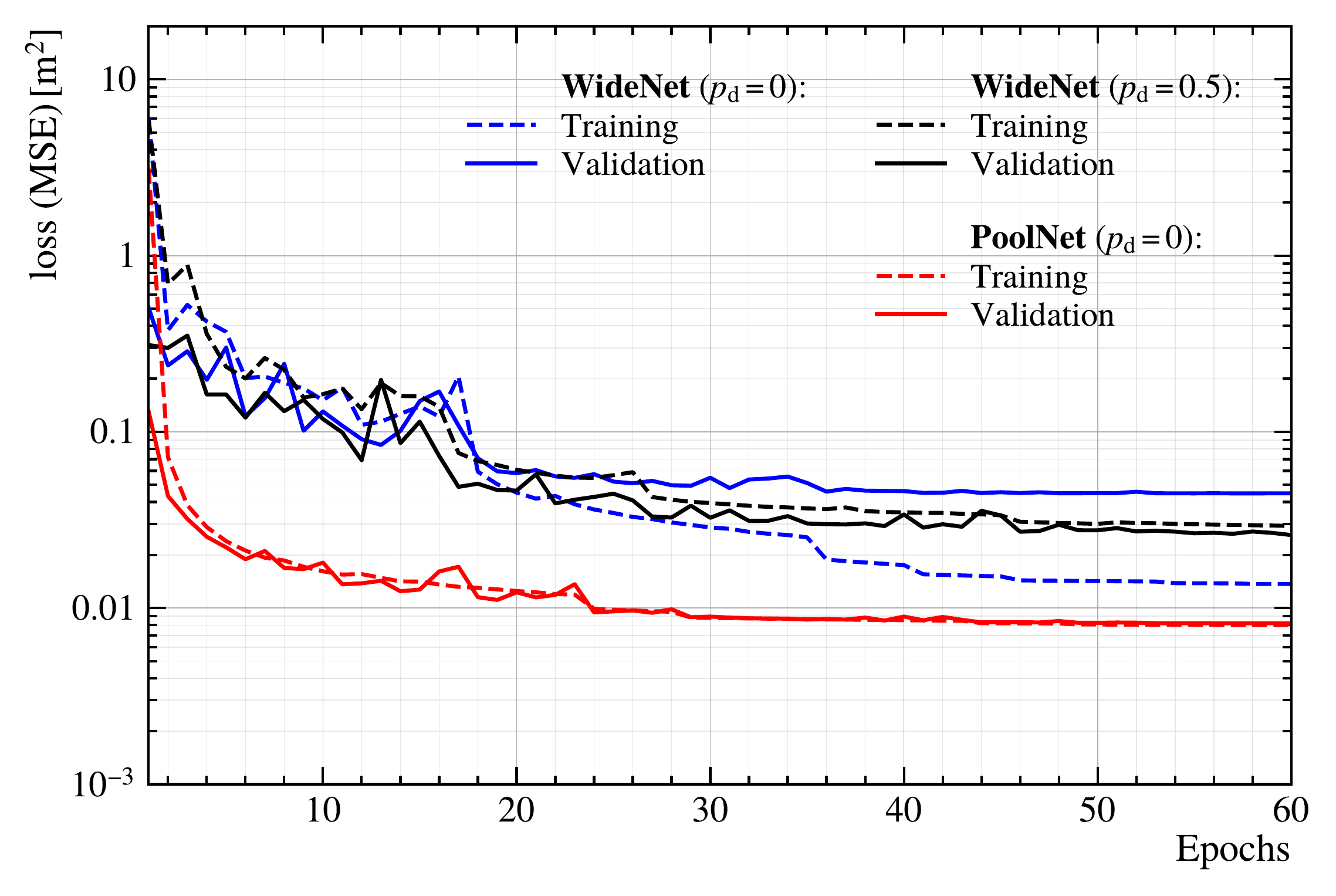}
        \caption{Training curves for \WN{} and \PN{}. The MSE-loss is shown for both networks against the trained epoch: blue for \WN{} without dropout, black for \WN{} with dropout (dropout ratio of last layer $p_{\mathrm{d}}=0.5$), and red for \PN. The validation loss is shown as a solid line for all cases, while the training loss is depicted with a dashed line.}
        \label{fig:pooling_training_1}
    \end{figure}

    We start to train \PN{} and \WN{} without regularization as explained in section \ref{sec:reco}.
    The training and validation losses are plotted against the epochs in Figure \ref{fig:pooling_training_1} for both networks.
    As expected \cite{CNN_goodfellow, CNN_geron2019hands}, training and validation losses improve comparatively for both networks during early epochs.
    However, stronger fluctuations occur for \WN{}, indicating a less stable training compared to \PN{}.
    The validation loss of \PN{} levels out almost an order of magnitude below the validation loss of \WN{}, at about \SI{8.2e-3}{\meter^2} compared to \SI{4.5e-2}{\meter^2}.
    \PN{} performs significantly better than the unregularized \WN{} in terms of MSE. 
    After the 29th epoch a separation of training and validation loss can be observed for \WN{} indicating that the network is overfitting \cite{CNN_goodfellow, CNN_geron2019hands}.
    Overfitting is not observed for \PN{}.
    As expected, \PN{} is less susceptible to overfitting due to the lower number of parameters.
    Additionally, its training is significantly faster.
    Training \WN{} takes about \SI{3.2}{} times longer per epoch compared to \PN.
    Here and in the following, we train all networks for \num{60} epochs since we observe no significant improvement afterward. 
    The validation losses are stable for at least five epochs as well.

\begin{figure}[htbp]
        \centering
        \includegraphics[width=\textwidth]{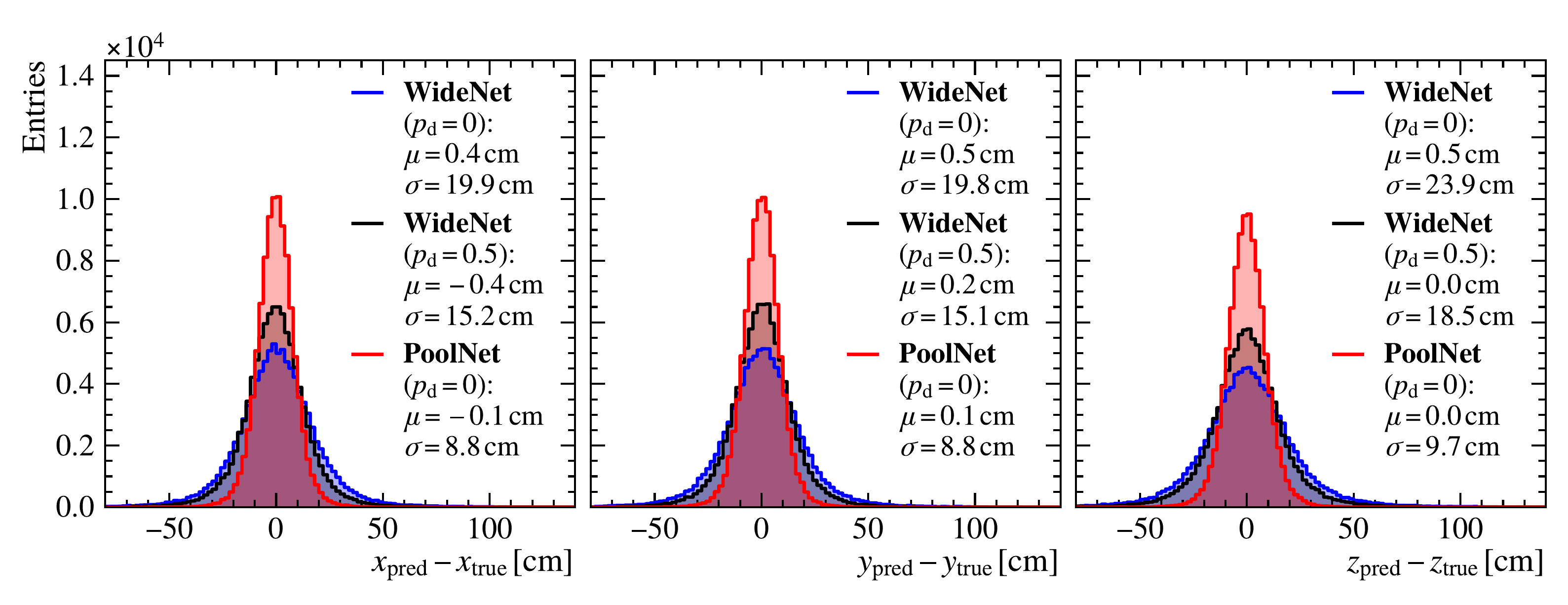}
        \caption{Vertex prediction deviation for \WN{} and \PN. The differences between network prediction and true values are plotted for $x,y,z$ evaluated on the test set: blue for \WN{} without dropout, black for \WN{} with dropout (dropout ratio of last layer $p_{\mathrm{d}}=0.5$), and red for \PN. The bias $\mu$ and resolution $\sigma$ (standard deviation) are given for all distributions. Their statistical uncertainties are \SI{0.32}{\percent} and \SI{0.22}{\percent} of the corresponding $\sigma$ respectively.}
        \label{fig:pooling_eval_1}
    \end{figure}

    For a more in-depth comparison, we evaluate the prediction uncertainty of each Cartesian coordinate on the test set.
    The parameter values from the epoch with the minimal validation loss is used for all network evaluations.
    We refer to the Bessel corrected sample standard deviation $\sigma$ as the resolution and the sample mean as the bias $\mu$.
    They are shown in Figure \ref{fig:pooling_eval_1} for all three coordinates and both networks.
    All distributions peak at about zero.
    The biases $\mu$ are small in comparison to the widths of the distributions and thus effectively do not contribute to the overall prediction uncertainty.
    As expected from the validations loss, the resolution of the \PN{} prediction is significantly better than that of \WN{}.
    For all three space dimensions \PN{} outperforms \WN{} by more than a factor of two in standard deviations $\sigma$.
    \WN{} has a resolution of about \SI{20}{\centi \meter} in $x$ and $y$ compared to \SI{8.8}{\centi \meter} for \PN{}.
    The resolutions in $z$ are \SI{23.9}{\centi \meter} for \WN{} and \SI{9.7}{\centi \meter} for \PN.
    
    As \WN{} starts overfitting for later epochs, the training does not yield the best performance for this network configuration.
    We train \WN{} again including a regularization via \emph{dropout} in the last layer.
    We use a \emph{dropout} ratio of $p_d = \SI{50}{\percent}$.
    All other training and network parameters are kept the same.
    The regularized \WN{} training is shown in Figure~\ref{fig:pooling_training_1} as well.
    The minimum validation loss improves by almost a factor of two, to about \SI{2.6e-2}{\meter^2}.
    Nevertheless, \PN{} still outperforms \WN{} by about a factor of four in terms of MSE.
    Overfitting is not observed anymore for the regularized \WN{}.
    The runtime of the training is not impacted by the \emph{dropout} regularization.

    \WN's resolution, including regularization, is shown in Figure \ref{fig:pooling_eval_1} as well.
    It can be seen that the resolution of the regularized \WN{} improves to about \SI{15.2}{\centi \meter} for $x$ and $y$ and to \SI{18.5}{\centi \meter} for $z$.
    As expected from the validation loss, \PN's resolution is still about a factor of two better.
    
    \subsection{Deep Network Performance}
    
\begin{figure}[htbp]
        \centering
        \includegraphics[clip, trim=2cm 6.5cm 1.cm 6.cm, width=0.95\textwidth]{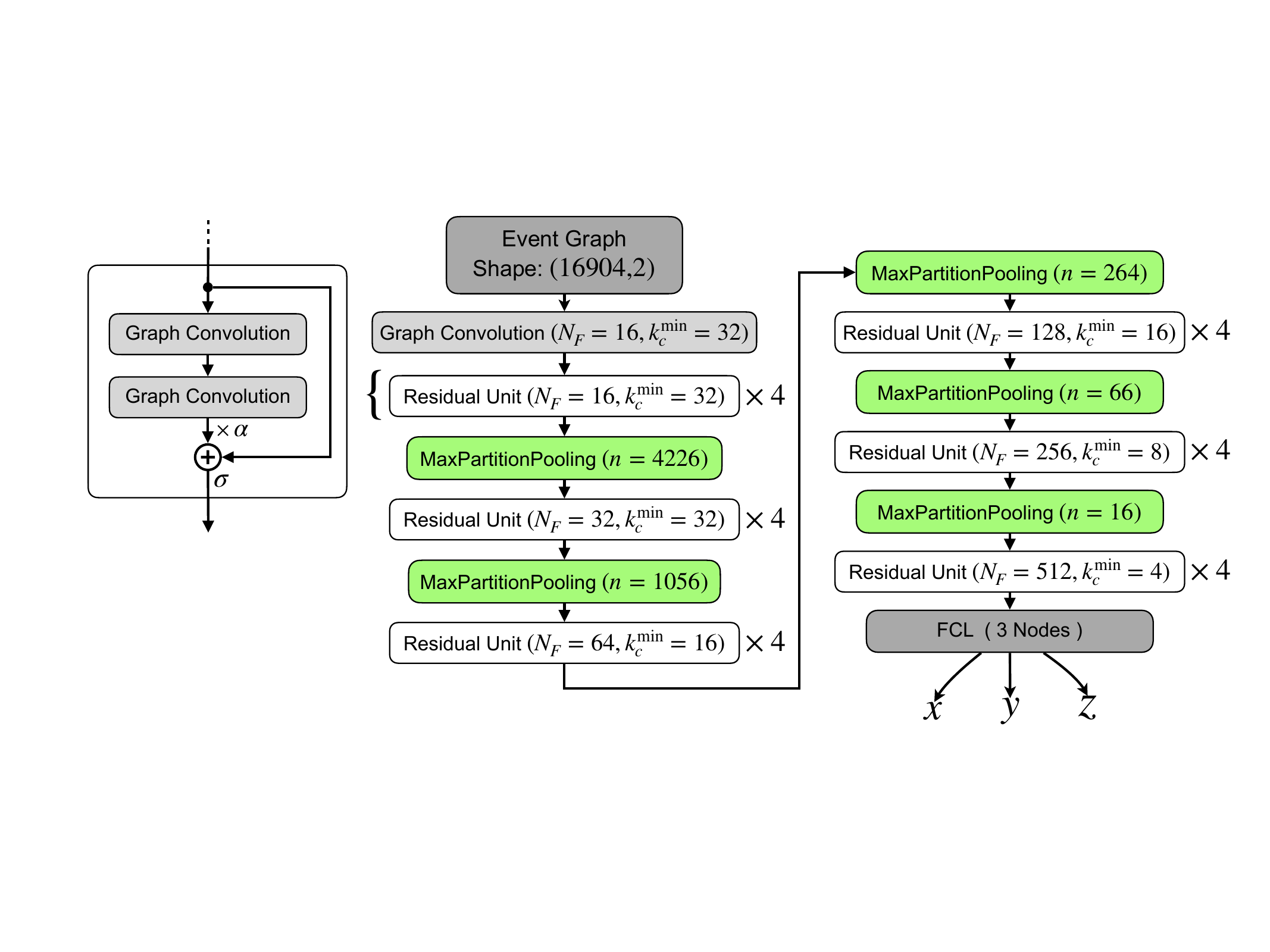}
        \caption{\DN{} structure. The network is fed with the node input. The first layer is a single graph convolution layer. All following convolution layers are arranged in twenty-four \emph{Residual} units \cite{ResNet} with additional residual weight \cite{ReZero}. For each convolution layer, the number of feature maps $N_F$, and the minimum connections per node $k_C^{\mathrm{min}}$ are given. After every fourth \emph{Residual} units, a partition pooling layer, highlighted in green, reduces the number of nodes $n$ by about a factor of four. The network output is the prediction of the three vertex coordinates $x,y,z$, implemented as a fully connected layer with linear activation.}
        \label{fig:deep_architecture}
    \end{figure}

    For real-life applications, computation time and memory are usually limited.
    The faster training times and reduced dimensionalities in CGNs that include pooling effectively permit deeper CGNs.
    We deepen \PN{} by doubling the amount of \emph{ResNet} blocks between the pooling layers, as depicted in Figure \ref{fig:deep_architecture}.
    The first four pooling layers are kept the same as in \PN{}, but we add a fifth layer that pools down to \num{16} nodes.
    Subsequently, this is followed by four additional \emph{ResNet} blocks.
    Additionally, we change the amount of feature maps for each \emph{ResNet} block to the typical increase of maps for later layers \cite{CNN_goodfellow, CNN_geron2019hands}, starting from \SI{16}{} for the first block to \SI{512}{} feature maps for the last one.
    We can not use this amount of feature maps for \WN{} due to resource limitations.
    As the number of nodes decreases after each pooling layer, we also reduce $k_C^{\mathrm{min}}$ from \SI{32}{} for the first GCN layer to \SI{8}{} for the last one.
    The new network consists of \num{2824203} trainable parameters in total, which is about a factor 1.7 more than \WN.
    A network with about twice as many parameters as the overfitting \WN{}, and additional partition pooling is reasonable for probing the benefits of pooling in large CGNs.
    We refer to the new network as \DN.
    
\begin{figure}[htbp]
        \centering
        \includegraphics[width=0.75\textwidth]{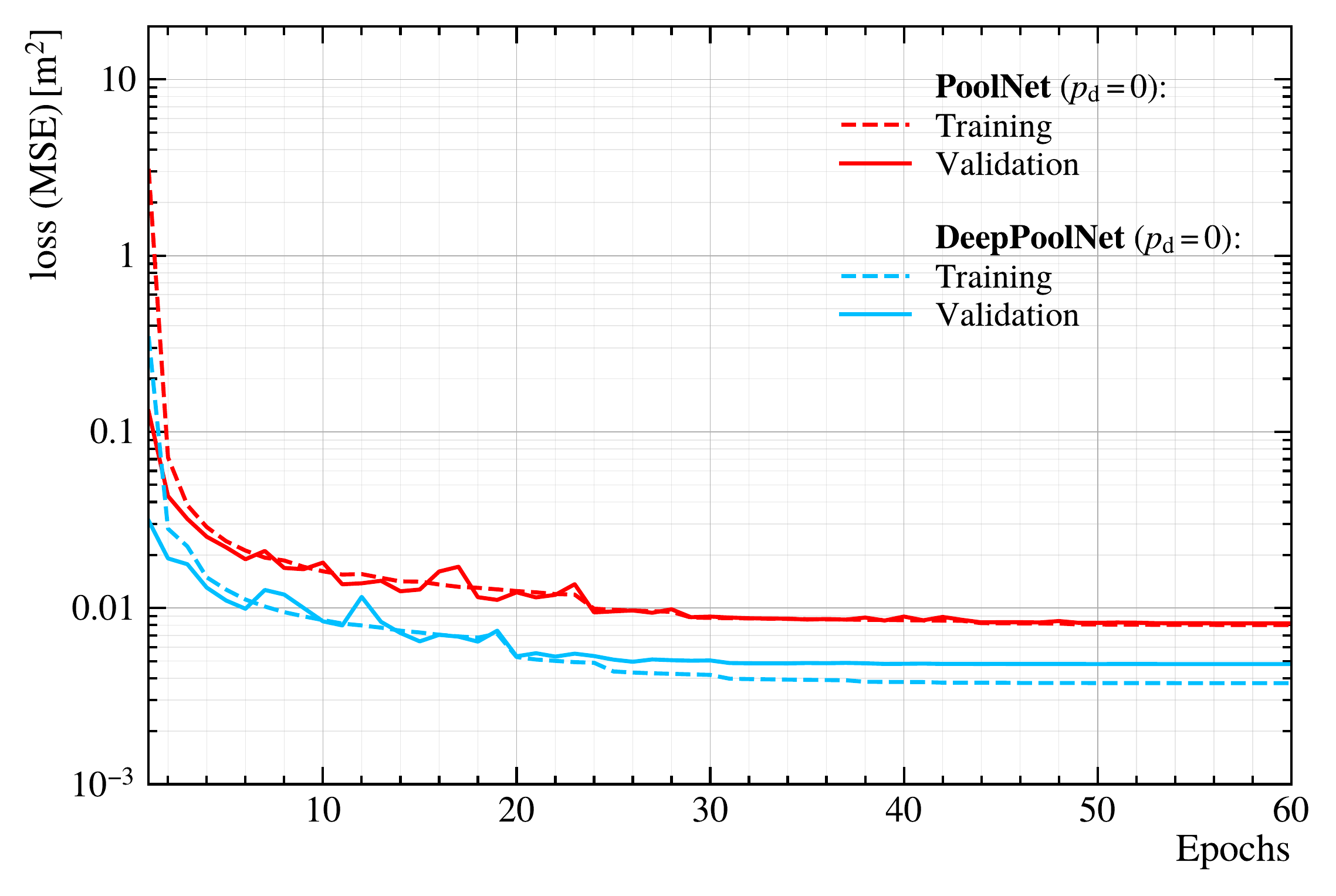}
        \caption{Training curves for \PN{} and \DN{}. The MSE-loss is shown for both networks against the trained epoch: red for \PN{} and sky blue for \DN. The validation loss is shown as a solid line for both cases, while the training loss is depicted with a dashed line. The dropout ratio $p_{\mathrm{d}}$ of the last layer is zero for both.}
        \label{fig:pooling_training_3}
    \end{figure}

\begin{figure}[htbp]
        \centering
        \includegraphics[width=0.99\textwidth]{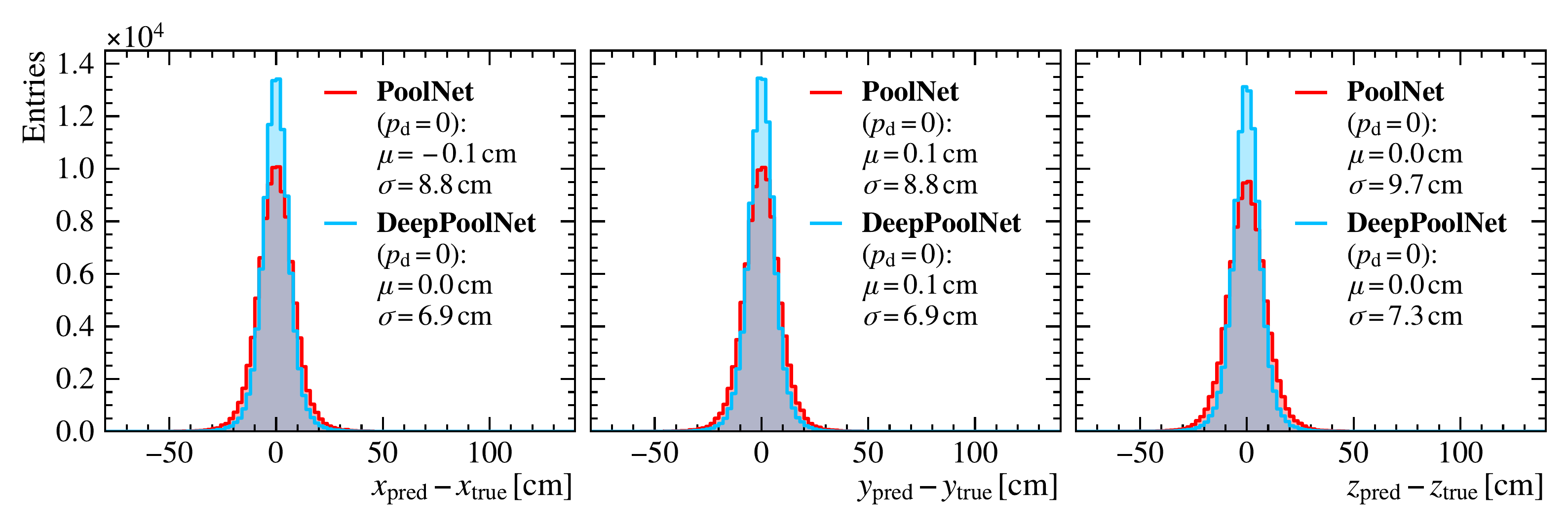}
        \caption{Vertex prediction deviation for \PN{} and \DN{}. The differences between network prediction and the true values are plotted for $x,y,z$ evaluated on the test set: red for \PN{} and sky blue for \DN. The bias $\mu$ and resolution $\sigma$ (standard deviation) are given for all distributions.
        Their statistical uncertainties are \SI{0.32}{\percent} and \SI{0.22}{\percent} of the corresponding $\sigma$ respectively.
        The dropout ratio $p_{\mathrm{d}}$ of the last layer is zero for both.}
        \label{fig:pooling_eval_3}
    \end{figure}

    The training curve is compared with the shallower \PN{} in Figure \ref{fig:pooling_training_3}.
    \DN's validation loss decreases faster than \PN's and stays significantly below it for all epochs.
    The minimum validation loss for \DN{} is about \SI{4.8e-3}{\meter^2}.
    Despite the large number of trainable parameters, overfitting is not observed for \DN, only a small generalization gap.
    We also compare the prediction resolutions on the test set in \mbox{Figure \ref{fig:pooling_eval_3}}.
    \DN{} achieves a resolution of \SI{6.9}{\centi \meter} in $x$ and $y$ and \SI{7.3}{\centi \meter} in $z$, which is significantly better than \PN.
    \DN, which can be realized only because of partition pooling, yields the best performance of the three networks tested.

    \section{Conclusion}
    We have presented a pooling scheme that uses partitioning to create non-overlapping dense pooling kernels for graph neural networks.
    The implementation of our algorithm for \emph{tensorflow} can be accessed in \cite{git_repo}.
    To demonstrate the applicability of partition pooling, we have simulated a typical neutrino detector and reconstructed the interaction vertices using GCNs with partition pooling.
    We have utilized \emph{K-means} partitioning, though other algorithms can be tested.
    The network's architecture has been inspired by established CNN architectures used for image recognition tasks.
    Comparing the pooling network with a similar CGN without pooling, we have shown that our method helps to improve the performance. 
    Partition pooling reduces the dimensionality for later layers, which leads to reduced computational resources required for training.
    This improvement is comparable to the improvement gained by including pooling in CNN for image recognition tasks~ \cite{image_pooling}.
    This has allowed us to create an even deeper CGN based on the same pooling chain, which is outperforming the shallower networks significantly.
    
    Partition pooling in CGNs can principally be evaluated for all reconstruction or classification tasks at any particle detector with a geometric sensor arrangement.
    Though, there might be more suitable partitioning algorithms than K-Means for detector geometries with particular symmetries.
    The optimal network architecture and number of partition layer has to be investigated for each task and with a corresponding hyperparameter optimization.
    Additionally, the specific detector systematics and biases need to be considered, e.g., by cross-checking the performance on calibration data.

    In conclusion, partition pooling makes deeper GCNs accessible.
    The faster training permits a more extensive hyperparameter optimization.
    Partition pooling is not restricted to our detector simulation but is universally applicable to similar graphs representing real particle detectors.
    It allows to adopt successful image processing architectures, e.g., \emph{VGGNet} \cite{vggnet}, \emph{ResNet} \cite{ResNet} and \emph{DenseNet} \cite{densenet}, for graph neural network applications in particle physics.
    More elaborate architectures like \emph{AutoEncoders} and \emph{GAN}s \cite{CNN_goodfellow, CNN_geron2019hands} can be adopted for CGNs using partition pooling as well.
    Summarized, the capabilities of partition pooling promise improved performances for GCN applications in particle physics.

\appendix
\section{Detector Simulation}
    \label{apd:toy_mc}
    Our detector setup is inspired by typical liquid scintillator detectors, common in neutrino physics for intermediate neutrino energies \cite{reactor_neutrino_book, reactor_neutrino_paper}.
    The core of these detectors is a vessel that contains organic liquid scintillator.
    The volume is instrumented by photomultiplier tubes (PMTs) situated outside the vessel, facing inwards through the transparent vessel.
    Neutrino interactions within the vessel excite the liquid scintillator, which subsequently deexcites by emitting photons isotropically \cite{birks}.
    As the track lengths of the exciting particles are small at moderate energies, the photons are emitted close to the interaction vertices.
    The events are considered to be point-like, thus they can be parameterized by their vertex positions and number of photons generated.
    After propagating through the detector medium, some of the emitted photons are detected by the PMTs.
    The intrinsic light yield of modern organic liquid scintillators is in the order of \SIrange{e3}{e4}{\photons/\MeV^{-1}} \cite{DoubleChooz:2006vya, JUNO:2015zny, borexino}.
    
    We use a rather large cylindrical setup of \SI{15}{\meter} radius and \SI{30}{\meter} height. It is read out by \num{16904} PMTs.
    This large number of sensors emphasizes the advantages of pooling. 
    It matches the complexity of the newest generations of detectors \cite{JUNO:2015zny, JUNO:2015sjr}.
    We model the PMTs as flat disks with a radius of \SI{20}{\centi \meter}.
    The PMTs on the barrel sides of the vessel are equally spaced on horizontal rings, as shown in Figure \ref{fig:toy_graph}.
    The rings are placed above each other without any rotation offset.
    PMTs of different rings are placed directly above each other.
    Additional PMTs are placed on rings above the lid and below the base.
    
    We have generated \num{e6} events uniformly distributed over the cylindrical volume.
    An example event is displayed in Figure \ref{fig:event_sample}.
    For each event, a random number of photons is created, uniformly distributed between \num{1000} and \num{10000}.
    The photons are emitted isotropically, and move along straight lines.
    For simplicity, we neglect scattering, but to add a small attenuation effect, we include absorption with a universal absorption length of \SI{15}{\meter}.
    Unless a photon is absorbed within the medium, we check if its ray intercepts with a PMT.
    A photon that intercepts a PMT is randomly registered by the PMT in \SI{30}{\percent} of the cases, which is a typical efficiency \cite{hamamatsu}.
    
    We assign an arrival time to each registered photon.
    In our simplified simulation, the measured arrival time includes three effects:
    
    i) As the liquid scintillator creates photons by deexcitation, the emission time $t_0$ for each photon is slightly different.
    $t_0$ is randomly drawn from a sum of three exponential distributions \cite{birks} for each photon 
    \begin{equation}
        t_0 \ \sim \quad \frac{f_1}{\tau_1} e^{-\frac{t}{\tau_1}} + \frac{f_2}{\tau_2} e^{-\frac{t}{\tau_2}} + \frac{f_3}{\tau_3} e^{-\frac{t}{\tau_3}}
        \ .
        \label{eq:scint}
    \end{equation}
    The values of the parameters can be taken from Table \ref{tab:ls_params}.
    
    ii) The time of flight $t_{\text{ToF}}$ until the photon hits the surface of the PMT is given by 
        \begin{equation}
        t_{\text{ToF}} = \frac{s \cdot n_r}{c_0} 
        \ ,
    \end{equation}
    where $s$ is  the traveled distance, $c_0$ the speed of light in vacuum,  and $n_r = \SI{1.53}{}$ the refractive index of the medium.
    
    iii) The arrival time is measured with limited precision by the PMT \cite{hamamatsu}.
    Therefore we smear the arrival time by a normal distribution $t_{\text{smear.}} \sim \mathcal{N}(\mu,\sigma) $ with $\mu=0 $ and $\sigma=\SI{1}{ns} $.
    
    The resulting arrival time for each registered photons is then given by
    \begin{equation}
        t_{\text{meas.}} =  t_0 +  t_{\text{ToF}}  +  t_{\text{smear.}}
    \end{equation} 
\mbox{}
    
    \begin{table}[htp]
        \centering
        \caption{Parameters for the simulation of the liquid scintillator time profile.}
        \begin{tabular}[htbp]{l|c|c|c|c|c|c}
         parameter   & $f_1$      & $\tau_1$              & $f_2$       & $\tau_2$               & $f_3$       & $\tau_3$                \\ \hline
         value      & \SI{0.9}{} & \SI{5}{\nano \second} & \SI{0.08}{} & \SI{50}{\nano \second} & \SI{0.02}{} & \SI{300}{\nano \second}
        \end{tabular}
        \label{tab:ls_params}
    \end{table}
    




\end{document}